\documentstyle[draft]{mn}

\input epsf
\input psfig.sty

\begin{document}

\newcommand{\kms}{\,{\rm km}\,{\rm s}^{-1}}
\newcommand{\mum}{$\,\mu$m}
\def\undertext#1{$\underline{\smash{\hbox{#1}}}$}
\def\IRAS{{\sl IRAS\/}}

\journal{Preprint UBC-COS-98-04, astro-ph/9810444}
\title[B\,1933+503, a dusty radio quasar at $z>2$]
{B\,1933+503, a dusty radio quasar at $\bmath{z>2}\,$:\\
implications for blank field sub-mm surveys?}

\author[S.C.~Chapman et al.]
       {Scott C.~Chapman,$^1$ Douglas Scott,$^1$ Geraint F.~Lewis,$^{2,3}$
	Colin Borys$^1$
\newauthor and Gregory G.\ Fahlman$^1$\\
        \vspace*{1mm}\\
        $^1$Department\ of Physics \& Astronomy,
        University of British Columbia,
        Vancouver, B.C.~V6T 1Z1,~~Canada\\
	$^2$Department\ of Physics \& Astronomy,
        University of Victoria,
        Victoria B.C.~V8W 3P6,~~Canada \\
	$^3$Department of Astronomy, Box 351580,
	University of Washington, 
	Seattle, WA~98195,~~U.S.A.}

\date{Accepted ... ;
      Received ... ;
      in original form ...}
\pubyear{1998}
\pagerange{000--000}
\maketitle

\begin{abstract}
We present a detailed mm-wave and optical study of the gravitational
lens system B\,1933+503, discovered by Sykes et al.~(1998) in the radio.
This object is probably the most complex
lens system known, with 10 lensed components within  a radius of one
arcsecond. It is potentially important as a probe of the Hubble constant,
although
no  optical counterpart has  thus far been observed  down to $I\,{=}\,24.2$.
We have obtained new sub-millimetre detections
at 450\mum, 850\mum\ and 1350\mum.  We have
also  constrained  the  possible  dust   emission  from  the  proposed
foreground lensing galaxy using a $K$-band adaptive optics image and 
CO(5--4) measurements.  A lensing model  is
constructed, taking the foreground elliptical galaxy at $z\,{=}\,0.755$ as the
lensing mass.  From this we derive a  scenario from  which to model the
sub-millimetre emission.  
Several arguments then point to the
source in the B\,1933+503 system lying above a redshift of 2. We speculate
that unlensed relatives of this source may constitute a sizable fraction
of the 850\mum\ source counts.
\end{abstract}


\newcommand{\ffffff}[1]{\mbox{$#1$}}
\newcommand{\scnd}{\mbox{\ffffff{''}\hskip-0.3em .}}
\newcommand{\scmd}{\mbox{\ffffff{''}}}
\newcommand{\minu}{\mbox{\ffffff{'}\hskip-0.3em .}}

\section{Introduction}

The sub-millimeter waveband has recently  become an invaluable tool for
investigation of  the properties of  high redshift galaxies. An early result
using the Sub-millimetre Common User Bolometer Array (SCUBA) by
Smail, Ivison and Blain (1997), showed  that a  much larger
population  of dusty, high redshift  galaxies  existed than previously
thought.  Since then, much  work  has gone  into identifying just what
types of galaxies  make  up this  sub-mm  bright population, and  what
mechanisms may be responsible for the rest-frame far-IR emission
\cite{Barger,Ealetal,Hugetal,Ivietal,Smaetal}.
An important application of SCUBA results is in constraining the global
star formation rate, and a related unresolved issue
is determining the fraction of the sub-mm emission
due to AGN activity, as opposed to a star-bursting process.

The recently  discovered radio object, B\,1933+503, is among the most
complex   gravitational lens  systems  yet  observed, with 10 lensed
components within one  square arcsecond \cite{Syketal}.  The existence
of  such a large number of  images makes this  a potentially important
candidate for determining the Hubble constant, through the measurement
of relative time  delays  between image  pairs.  However Hubble  Space
Telescope $V$ and $I$ band images of B\,1933+503 (also described in Sykes et
al.~1998) reveal the   lensing    galaxy, but  no   apparent   optical
counterpart for the  source down to about $I\,{=}\,24$.
Hence the  source redshift has  not yet
been  obtained.  There are two  scenarios  which could account for the
lack  of optical emission: 1) the source object is intrinsically
underluminous in the optical;  or 2) the source or foreground lens is
heavily dust enshrouded  and thereby obscured at optical  wavelengths.

In order to  characterise this  object in  the framework of  dusty
high redshift sources,  and  perhaps to  shed   some light on  the source
redshift, we  obtained sub-millimetre   continuum photometry
at 450\mum,  850\mum\ and 1350\mum\ using the
SCUBA instrument on the James  Clerk Maxwell Telescope.
We have also indirectly constrained the dust obscuration in the lensing
galaxy using an upper limit to  CO line emission, and  a high  resolution,
adaptive optics $K$~band image.
Using the combined multi-wavelength data-set, we
attempt to further constrain the properties of this system, and towards this
end have also developed a new lensing model.

%
%
\setcounter{table}{0}
\begin{table*}
\begin{center}
\caption{The observed properties of B\,1933+503, and the foreground
elliptical, from the radio 
to the optical waveband.  None of the source quantities are corrected
for the effects of gravitational lensing, which has a likely amplification
factor of $\sim14$ in the radio and perhaps 12 in the sub-mm and
far-IR.\hfil }
\begin{tabular}{lcccl}
\noalign{\medskip}
\noalign{\smallskip}
Property & Telescope & B\,1933$+$503& {Comment} \cr
\hline
\noalign{\medskip}
\undertext{Distant AGN (source):}&&&\cr
\noalign{\smallskip}
$\alpha\,$(J2000) & & $19^{\rm h} 34^{\rm m}$30\fs899 &
from VLA map component 4 \cite{Syketal} \cr
$\delta\,$(J2000) & & $50^{\circ} 25^\prime$23\farcs22 & \cr
\noalign{\medskip}
Redshift estimate from lens model  & & $>2.34 $ &
 see Section~\ref{sec:lensing}\cr
\noalign{\medskip}
Flux density at: & & & \cr
\noalign{\smallskip}
$\>\>1.7$\,GHz & VLA (D)& $75.9\,$mJy &  All radio components
 \cite{Syketal}\cr
$\>\>5$\,GHz & VLA (D)& $57.5\,$mJy &  All radio components
 \cite{Syketal}\cr
$\>\>8.4$\,GHz & VLA (D)& $41.1\,$mJy &  All radio components
 \cite{Syketal}\cr
$\>\>15$\,GHz & VLA (D)& $37.1\,$mJy &  All radio components
 \cite{Syketal}\cr
\noalign{\smallskip}
$\>\>1350\,\mu$m & JCMT & $30 \pm 7$\,mJy & SCUBA photometry \cr
$\>\>850\,\mu$m & JCMT & $24.1 \pm 2.6$\,mJy & SCUBA photometry \cr
$\>\>450\,\mu$m & JCMT & $114 \pm 17$\,mJy & SCUBA photometry \cr
\noalign{\smallskip}
$\>\>100\,\mu$m & \IRAS & $<443$\,mJy & {\tt XSCANPI} limit\cr
$\>\>60\,\mu$m & \IRAS & $<136$\,mJy & {\tt XSCANPI} limit\cr
$\>\>25\,\mu$m & \IRAS & $<69$\,mJy & {\tt XSCANPI} limit \cr
$\>\>12\,\mu$m & \IRAS & $<69$\,mJy & {\tt XSCANPI} limit \cr
\noalign{\smallskip}
$\>\>K$ & CFHT/AOB & $\ga20.3$ &  Point source limit \cr
$\>\>I$ & HST & $\ga24.2$ &  Point source limit \cite{Syketal} \cr
$\>\>V$ & HST & $\ga25.1$ &  Point source limit \cr
\noalign{\medskip}
\undertext{Foreground elliptical galaxy (lens):}&&&\cr
\noalign{\smallskip}
$\>\>K_{\rm ap}$ & CFHT/AOB & $17.1\pm 0.5$ &  Aperture magnitude \cr
$\>\>I_{\rm ap}$ & HST & $20.6\pm 0.2$ &  Aperture magnitude
 \cite{Syketal}\cr
$\>\>V_{\rm ap}$ & HST & $>22.5$ &  Aperture magnitude limit
 \cite{Syketal}\cr
$\>\> {}^{12}$CO(5--4) & JCMT & ${<}\,490\,{\rm mK}\kms$ &  Assuming
$300\kms$ linewidth \cr
$\>\>$Redshift& Keck& $0.755\pm 0.001$ & C. Fassnacht (private communication)
\cr
\end{tabular}
\end{center}
\end{table*}

The paper is organized  as  follows.  We  first describe the  new  and
previously  existing  observations of B\,1933+503.  We  then  describe a new
lensing model using the foreground  elliptical galaxy at $z\,{=}\,0.755$ as the
lensing mass.  Various emission mechanisms are explored for the sub-mm
flux, taking   into  account the   flat-spectrum radio components,
dust, and the  possible  contamination  of  the signal by   foreground
emission in the   $z\,{=}\,0.755$ elliptical.  From this  we  derive the  most
likely scenario for this  object,  and consider why   it is  not  more
readily observable in the optical and near-infrared bands.

\section{Observations}

B\,1933+503 (hereafter `B\,1933') was discovered as 
part of the  Cosmic Lens All-Sky Survey~\cite{Jacetal},
and has already been investigated in some detail at radio and
optical wavelengths (see Sykes et al.~1998).  We summarize  the
existing data, as well as our new observations, for both B\,1933 and
the proposed lensing galaxy in Table~1.  We include the radio data,
as well as upper limits from \IRAS, derived  using the {\tt XSCANPI} application
provided by IPAC. These \IRAS\ limits are the 90 per cent confidence limits
for the flux per beam at the position of B\,1933. 

\subsection{Submillimeter observations}

Our observations   were
conducted with the  SCUBA instrument
\cite{Holetal} on the James Clerk Maxwell Telescope.
On 1997 December  3, we operated the  91 element Short-wave
array at 450\mum\ and the 37  element Long-wave array at 850\mum\
simultaneously in photometric mode, and also the single photometry pixel at
1350\mum,  giving half-power beam  widths of 7.5, 14.7, and 21 arcsec
respectively.    At   450/850\mum\,   five   `scans'  of 900  seconds,
consisting of 50 integrations each were taken, for a total integration
time of  1.25 hours.   The central pixel  of SCUBA   was fixed on  the
source, B\,1933.    The   1350\mum\  observation  consisted  of   an
additional integration time of one hour.
In  all cases, a 9-point jiggle pattern
was employed to reduce the impact of  pointing errors by averaging the
source signal over a slightly larger  area than the beam, resulting in
greater photometric  accuracy.    Whilst jiggling, the  secondary  was
chopped at 7.8125\,Hz by  90 arcsec in azimuth,  thus the other pixels
rotate  relative to the   sky through the  period  of the integration,
smearing out the effect of possible chopping on to other faint sources.
Pointing was checked hourly on the blazar 2036--419  and a sky-dip was
performed between each  15  minute scan   to  measure the  atmospheric
opacity.  The rms pointing errors  were below 2 arcseconds, while  the
average   atmospheric zenith  opacities   at   450\mum\, 850\mum\  and
1350\mum\ were fairly stable  with  $\tau$ being  0.51, 0.12 and  0.04
respectively.  However, there were  some short time-scale  variations,
presumably due to water vapour pockets blowing  over at high altitude,
which caused some parts of the data-set to be noisier (see Borys,
Chapman \& Scott~1998 for more details).

%
%
\setcounter{figure}{0}
\begin{figure}
\label{fig:cfht}
\begin{center}
\psfig{file=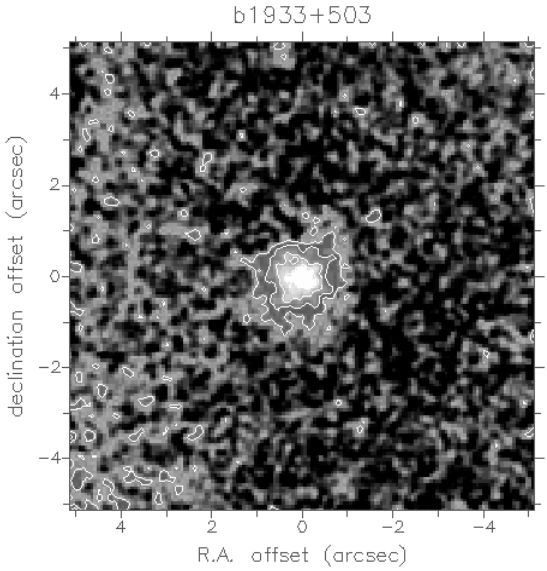,height=8cm,angle=0}
\caption{CFHT $K$-band adaptive optics image
of the 10 arcsec $\times$ 10 arcsec
field surrounding B\,1933, smoothed with a 0.31 arcsec
boxcar filter (corresponding to the resolution).
There is no indication of a central point source.}
\end{center}
\end{figure}

%
%
\begin{figure}
\label{fig:hst}
\begin{center}
\psfig{file=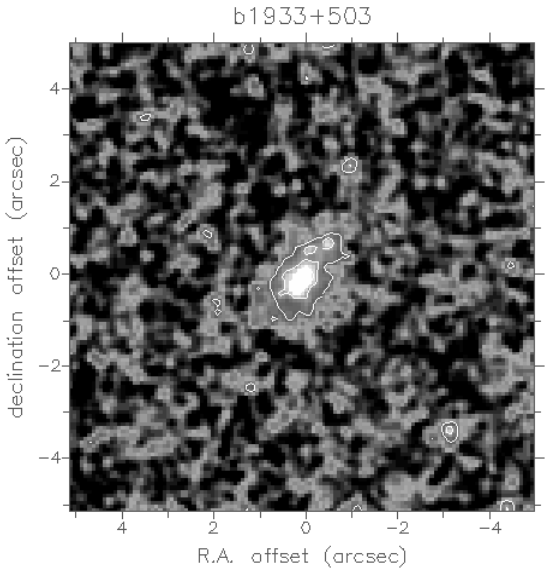,height=8cm,angle=0}
\caption{HST WFPC2 F814W ($I$-band) image, showing the same region as
above, also smoothed to 0.31 arcsec resolution.}
\end{center}
\end{figure}

%
%
\begin{figure}
\begin{center}
\psfig{file=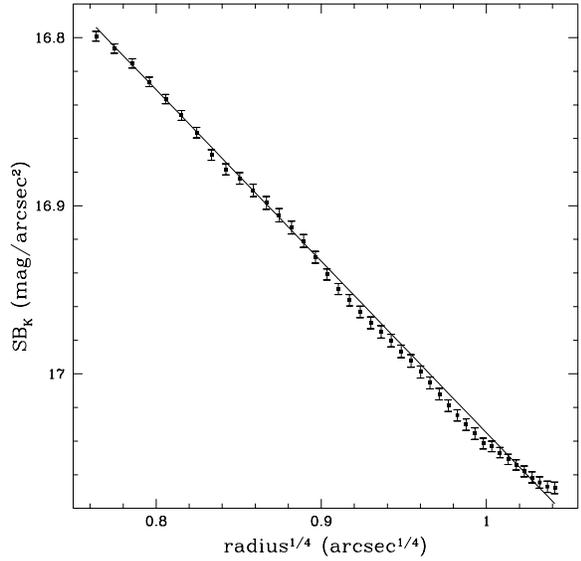,height=8cm}
\psfig{file=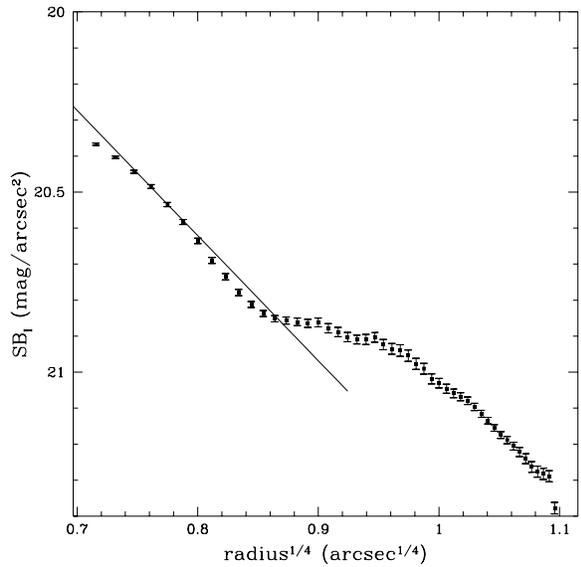,height=8cm}
\end{center}
\caption{Surface brightness profiles of the CFHT/AOB $K$-band (top) and
HST $I$-band (bottom) images,
smoothed with a 0.31 arcsec boxcar filter, and with an $r^{1/4}$ law
fit to the central regions.}
\label{fig:profiles}
\end{figure}

The data were reduced using both the  Starlink package SURF (Scuba User
Reduction Facility, Jenness \& Lightfoot~1998),
and independently using our own  routines
\cite{BorCS}.  Spikes  were  first carefully rejected  from the double
difference data. The data were then  corrected for atmospheric opacity
and sky subtracted using the median of all the array pixels except for
obviously bad pixels and the central pixels (the 1350\mum\ pixel currently
has no provision  for   subtracting sky  variations using   the  other
wavelength pixels).  The data were  then calibrated against Saturn  and
the compact {H}{II}
region  K3--50,  which were also observed  during the
same  observing shift. The two 850\mum\  and 1350\mum\ calibrations agreed
with each  other and also with the  standard gains to within 10 per cent.
However,
at 450\mum, K3--50    is extended and  variable,  and  is  not a  good
calibration source, while the  Saturn 450\mum\ calibration agreed with
standard gains to within 25 per cent.

We also  used   the  JCMT  RxB3 $345\,$GHz  band   heterodyne
receiver on 1998 January  3, with   the  widest   bandwidth possible
($920\,$MHz)   to    observe  the   $^{12}$CO(5--4)   $575\,$GHz line,
redshifted to $327.64\,$GHz. The intention was to check for a possible
high gas  mass from the lensing  galaxy.  No detection was apparent in
the data, which were binned to $38\,$MHz channels, with
a resulting rms noise of around $4\,$mK. We estimate an upper limit of
$0.49\,{\rm K}\,\kms$
for a Gaussian  line  with  $300\kms$  FWHM centred on  the
measured redshift $z\,{=}\,0.755$ (see section~\ref{sec:elliptical}).

\subsection{CFHT Adaptive Optics, $\bmath{K}$-band image}

We observed B\,1933 with
the Canada-France-Hawaii Telescope in  March 1998 with 
the  Adaptive Optics Bonnette (AOB, Rigaut et al.~1998)  in
place feeding  the   KIR  near-IR camera.  The detector has
$1024\times1024$ pixels at a  scale of $0.34$ arcsec per pixel
to Nyquist sample the
$J$-band  under  optimal observing conditions,   for   a field size   of
${\sim}\,36$ arcsec.
The average natural seeing throughout the
observations was worse than median for the site at $0.86$ arcsec, and the
performance  of the system was  therefore not optimal, with fairly low
Strehl  ratios.  The estimated   resolution   in the image near    the
$V\,{=}\,13.2$ guide star   is $0.22$ arcsec corresponding  to  a physical
scale  of  $890\,{\rm h}^{-1}$pc  at  $z\,{=}\,0.755$ (assuming
$\Omega_0=1$ and with $H_0=100h\kms\,{\rm Mpc}^{-1}$).
The  adaptive optics  corrected Point Spread Function (PSF)
degrades  in a predictable way  with distance from the guiding source,
as  a  function of  guide   star brightness,  atmospheric quality, and
wavelength of observation.  Therefore the PSF  for the position on the
detector of  B\,1933 was constructed from the  guide  star using a model
describing the PSF variation   over the observed field
\cite{Hutchings}.
B\,1933 lies $\sim15$ arcsec from the guide  star and the PSF
has a FWHM of 0.31 arcsec.

Flux and  PSF calibrations were  performed  using  the UKIRT  standard
stars fs13 and fs25 \cite{CasHaw}. 
Flat-field images were taken on the dome with the
lamps turned  on and   off to  account  for  the thermal glow   of the
telescope.   The image was processed   using  an algorithm described  in
Chapman  et   al.~(1998, in preparation),
where the  effects   of a noisy, variable  sky at
2.2\mum\ are reduced using  a high  signal to noise  thermal
image consisting   of all the  $K$-band  sky data taken  over a  5 night
observing run.  The signal to noise ($\sim 5$) of the resulting image 
is too low to merit deconvolution  techniques.  The image is
displayed, smoothed with a  0.31 arcsec boxcar filter, in
Fig.~1,
and the HST archive  WFPC2 $I$-band image in Fig.~2.
There is no
obvious detection of the  4 lensed point  sources of the radio core in
either  image, as discussed in the next section.
Furthermore, even the  foreground galaxy  is not  detected at $V$-band
with HST.

\section{Modelling}

\subsection{The $\bmath{z\,=\,0.755}$ elliptical} 
\label{sec:elliptical}
Before   considering  the  radio
source  itself, we examine  the  possibilities  for the  sub-millimeter
emission to arise in the $z\,{=}\,0.755$  foreground galaxy.  The [$I\,{-}\,K$]
colour of
the galaxy is  $3.5 \pm 0.5$, and [$V\,{-}\,I]>1.9$, which is fairly normal
for late-type galaxies at this magnitude and  redshift range (e.g.~Cowie
et al.~1994).
The colors are certainly consistent  with some dust reddening
though, with amplitude depending on how  red the $V\,{-}\,I$ index is.
We fit profiles to both the
HST  $I$-band   and CFHT $K$-band images   of  this galaxy, as shown in
Fig.~\ref{fig:profiles}. An
$r^{1/4}$ profile is a good  fit to the  core in both cases, with no
indication of a central cusp, i.e.~no detection of the radio source.
The HST $I$-band
image shows unresolved knots to the northwest, extending along what 
appears to be a spiral arm or tidal tail. 
The profile is distorted at this 
radius and is not well fit by a disc or bulge model. There is
some evidence for elongation   along   ${\rm PA}\,{=}\,-45^\circ$
of the $K$-band
image, especially in the first 500 second integration when the seeing
was  the best. This is consistent  with the HST  image, although the $K$
isophotes are closer to circular.

If we suppose that  all the sub-mm  emission comes from
this  galaxy instead of the radio source, we  can   calculate  the dust mass
required to produce this flux level (following e.g.~Hughes, Dunlop
\& Rawlings~1997).  The implied
dust mass is $M_{\rm d} \simeq 2\times10^8 h^{-2}{\rm M}_\odot$, with an
associated star formation rate of certainly
${>}\,500 {\rm M}_\odot\,{\rm yr}^{-1}$. This is a copious
amount  of dust, exceeding   that inferred for 
some ultra-luminous infrared galaxies
(e.g.~Eales and Edmunds 1996).  For  a fairly normal looking galaxy,
such a large star formation rate would be
difficult to  explain.  In addition, we shall see
in section 3.3 that the model fits to the spectral energy distribution
(SED) are difficult to reconcile if the sub-mm emission is at $z\,{=}\,0.755$.

We can also   use  the non-detection of  CO   at the redshift  of  the
elliptical galaxy to   estimate the maximum dust  mass  which could be
present.  Following Braine \& Dupraz (1994),  we can take the RMS level
achieved at a particular binning scale, and estimate the $1\sigma$
line intensity limit for lines of a given width.  For $300\kms$ FWHM lines
we find that $I_{\rm CO}\,{<}\,0.49\,{\rm K}\,\kms$, or
${<}\,14\,{\rm Jy}\kms$ for the JCMT at this frequency.
The  implied molecular  hydrogen gas mass
$M_{{\rm H}_2}\,{\la}\,6\times10^{9} h^{-2}{\rm M}_\odot$,
taking CO(5--4)/CO(3--3)$\,{\simeq}\,1$ and using a standard
$M_{{\rm H}_2}$ to $L_{\rm CO}$ conversion (e.g.~Barvainis et al.~1997).  An
accurate calculation of the mass of molecular gas would require detailed
modelling and multiple line constraints. However, if we take this value,
together with a typical value for
the gas to dust mass ratio, say $\sim500$, then the dust mass is
${\la}\,10^{7}h^{-2}{\rm M}_\odot$.  Even taking the modelling
uncertainties into
account, this is still an order of magnitude less than the 850\mum\
continuum measurement would  imply.  Uncertainties in the redshift and
linewidth probably alos reduce  this limit somewhat,   but these effects
still make it hard for the sub-mm flux to be coming from the elliptical
galaxy.  The  redshift would  have to be   in  error by around
0.005, which is much larger than the estimated error, before our detection
threshold would be much affected.
Moreover, the dust mass depends on linewidth only as
$M_{\rm d}\propto l^{1/2}$, which is a relatively
weak dependence.  Moreover, even assuming a dust to gas ratio as small
as $\sim100$,  the  implied  dust is still smaller
than that implied by the 850\mum\ flux.

Although the distorted, knotty morphology at $I$-band is suggestive of ongoing
star formation,  we can conclude from the above arguments that at most  
$\sim10$ per cent of  the sub-mm  emission could be
arising in the foreground elliptical.
This does not imply that the foreground lens cannot be a source
of optical obscuration. Indeed a class of `dusty lenses' has been discussed
in the literature \cite{Laretal,Lawetal,MalRT} where the background source 
is extremely reddened by small amounts of dust in the foreground lens.  
However, in at least one of these cases (MG1131+0456) the dusty lens hypothesis
has been questioned after {\sl HST}-NICMOS imaging of the source was obtained 
at $H$-band \cite{Kocetal}. Without such a detection of our source, we are
not able to rule out the possibility of some reddening by the foreground lens.
Therefore the obscuration of the optical source is currently an unknown
combination of self-absorption and absorption through the lens.

\subsection{Gravitational lensing}
\label{sec:lensing}

Nair  (1998) modelled the  lensing   of  B\,1933 using a   potential
derived from  the shape of   the HST $I$-band  image of  the  foreground
galaxy,   and the  positions of    the  detected radio  sources.    To
accurately interpret the   spectral energy distribution  of B\,1933,  we
first expanded on the Nair model using our new data, and then 
re-modelled the  system to assess  the most probable  redshift for the
source and the sub-mm to far-IR lensing amplification.  We will give a
more detailed analysis of the lens remodeling and
the implications for time delays elsewhere (Lewis et al.~in preparation).  

\subsubsection{Physical parameters of the lens/source}
To estimate the source redshift, the crucial parameter to evaluate is the 
model mass-to-light ratio, 
$\Upsilon_o = r\,({\cal M}/{\cal L})_o$, which contains the
quantity $r = D_{\rm s}/D_{\rm l-s}$, which
is the ratio of angular size distances from the observer to
the source and from the lens to the source.
Here ${\cal M}$ and ${\cal L}$
are the mass and luminosity of the galaxy within some aperature.
A constraint on $r = \Upsilon_{\rm c}/\Upsilon_{\rm o}$,
where $\Upsilon_{\rm c}$ is the expected
mass-to-light ratio of a normal elliptical galaxy, can be used to constrain the
redshift.

We write $\Upsilon_{\rm c}=\Upsilon_{\rm p}/f_{\rm evol}$, where
$\Upsilon_{\rm p}$ is the value appropriate
for present day ellipticals and $f_{\rm evol}$ is the evolutionary
correction accounting
for the fact that galaxies were brighter in the past. We adopt a blue-light
$\Upsilon_{\rm p}=(11.86 \pm 0.5)h$ from van der Marel (1991) and note that the
error is believed to be underestimated by about a factor of 3 due to the use
of a fundamental plane relationship to determine distances. The evolutionary
correction adopted is $f_{\rm evol}=2.5 \pm 0.5$ ($-1 \pm 0.2$  magnitudes)
from the models described in Pozzetti et al.~(1996).

The mass of the lensing galaxy is
calculated out to a major-axis distance of 1\,arcsec
from the Nair model, and the luminosity is obtained from the
reported photometry \cite{Syketal} within a
$2\,{\rm arcsec}\times1\,{\rm arcsec}$
aperature (which is slightly flatter
than the model axial ratio of  $b/a = 0.59$).
The apparent blue magnitude of the lensing galaxy is calculated from the
expression in Lilly et al.~(1996)
which relates the observed $I$-magnitude to the
blue magnitude using a K-correction derived from the CFRS data. The uncertainty
in this estimate is dominated by the photometric errors ($\sim 20$ per cent).

Two cosmological models, both with $\Lambda=0$, were considered:
(i) a standard inflationary model with $\Omega_0=1$ and (ii)
an open model with $\Omega_0=0.3$. The results are independent of the Hubble
constant and turn out not to be very sensitive to the cosmological
model parameters. However, the derived source redshift is $\em very$
sensitive to the parameters involved in the calculation of
$\Upsilon_{\rm c}$ and $\Upsilon_{\rm o}$. Using
the adopted mean values, we find $r=1.452$ and 1.640 in cases (i) and (ii)
respectively. These are close to the critical values corresponding to an
infinite source redshift and, formally, we find
$z_{\rm s} \simeq 20$ in either case.
>From a consideration of the uncertainties associated with
the various parameters that enter into the calculation, we suggest that a 
realistic error is $\Delta r/r \simeq 50$ per cent.
Taking the upper limit on $r$,
we obtain lower bounds of $z_{\rm s}=2.34$ and 2.18
for the two cosmological models.
A model SED is plotted in Fig.~\ref{fig:sed}b for this redshift lower limit
(for the $\Omega_0=1$ model).

There is a separate statistical argument about the source redshift, based
simply on the lensing optical depth.  Normally this is phrased in terms
of the probability of finding the lens at some redshift, given a source
redshift, but we can easily turn it around.
Kochanek (1993) has discussed the gravitational lens statistics for
isothermal spheres, showing, for example, that for a source at
redshift $z_{\rm s}$, the most probable location
of the lens lies at half the proper motion distance to the source.
For $z_{\rm l}=0.755$, and with a
uniform prior distribution on the redshift, we find a 95 per cent
confidence lower limit of $z_{\rm s}>1.1$, and a most likely redshift
of 2.9 (for an $\Omega_0=1$ model).  While not particularly constraining,
this lends some additional support to the above argument
for the source being at $z\ga2$.

\subsubsection{Re-modelling of the lens system}
We employed a simple elliptical potential
structure \cite{KocBLN}, and while the resulting  mass
profile is different from that used by Nair (1998), and we only used the
observed positions as a constraint, the model recovers  the
image positions to similar accuracy.  The total amplification of the
core  radio component in this  model  is 13.9, in reasonable agreement
with the 14.63 found by Nair (1998).

The degree to which the  flux from a  source is enhanced by the action
of gravitational lensing is dependent on the scale-size of the source,
leading to   possibly   pronounced differential magnification  effects
(e.g.~Schneider \& Weiss~1992).
We model the sub-mm emission as coming from a
dusty disc of $\sim200h^{-1}$  parsecs in extent, lying  perpendicular
to the axis  defined  by the  radio   core and  jet/lobes, and
assuming $\Omega_0=1$ to convert lengths to angles.
This corresponds to an  extended obscuring torus possibly related to the 
small-scale (few pc) dusty torus of the
standard unified  AGN  model (see for example the HST image of  NGC4261
from Jaffe et al.~1998).  Other sizes and geometries could of course be
considered; we only present this as one possibility.

\begin{figure*}
\begin{center}
\psfig{file=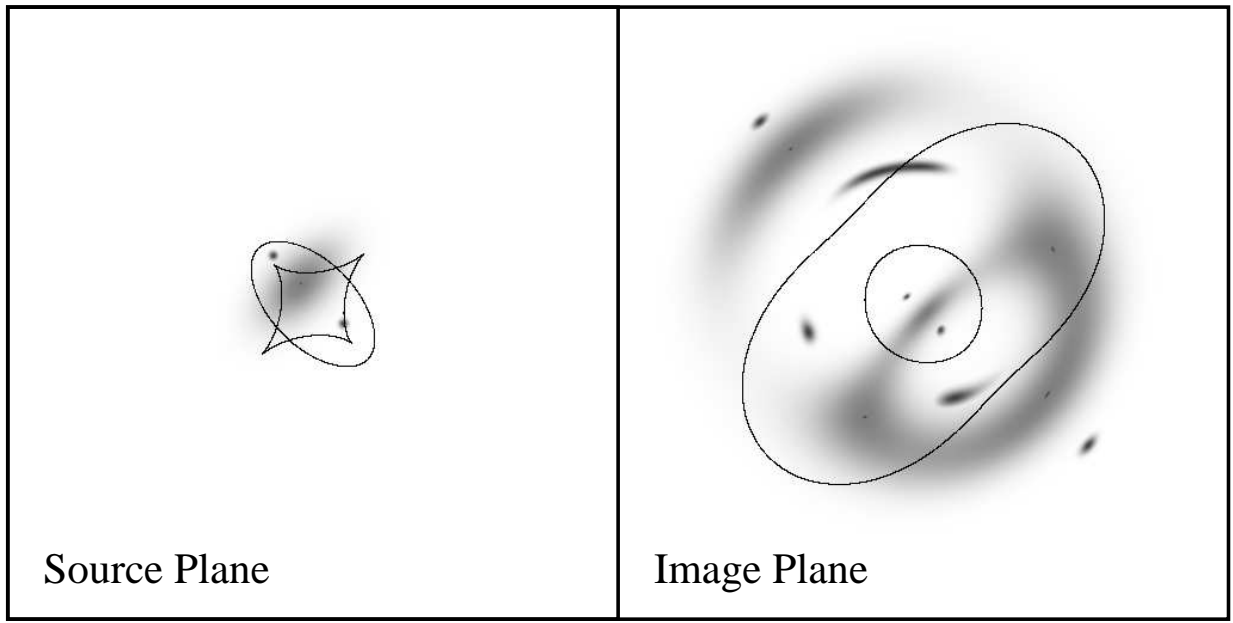,height=8.25cm,angle=270}
\end{center}
\caption{The result of our lens modelling for one particular example of a
dusty disc.
The left panel represents   the source plane,   while the right
panel  is the (observed)  image  plane. The solid lines denote
caustics in the source plane and critical lines in the image plane.
Each panel is 1.4 arcsec on a
side, and is oriented to align with fig.~1  of Sykes et
al.~(1998).  Darker shading represents both the shallow spectrum
(compact source) and shallow spectrum (extended `lobes')
radio emission, while lighter grey shows a model of the far-infrared and
sub-mm emission from a possible dusty torus.
The resulting image configuration can be compared to the
observed system.}
\label{fig:lens}
\end{figure*}

Fig.~\ref{fig:lens}
presents the results  of our modelling,
showing the source plane (left-hand panel) and the
resulting image plane  (right-hand panel).  The darker shading represents both
the steep and shallow spectrum radio emission (the central compact source
is a little hard to see in the source plane).
The field is 1.4 arcsec
in  extent and the  image  configuration can be  compared  directly to
fig.~1 in Sykes  et al.~(1998).  Lighter grey here represents the
sub-mm emission from  a dusty disc  model as described above.  The
resulting far-infrared and sub-mm amplification ($A_{\rm IR}$)
is 11.9.

%
%

\subsection{A dusty radio quasar}

\begin{figure}
\begin{center}
\leavevmode \epsfysize=8cm \epsfbox{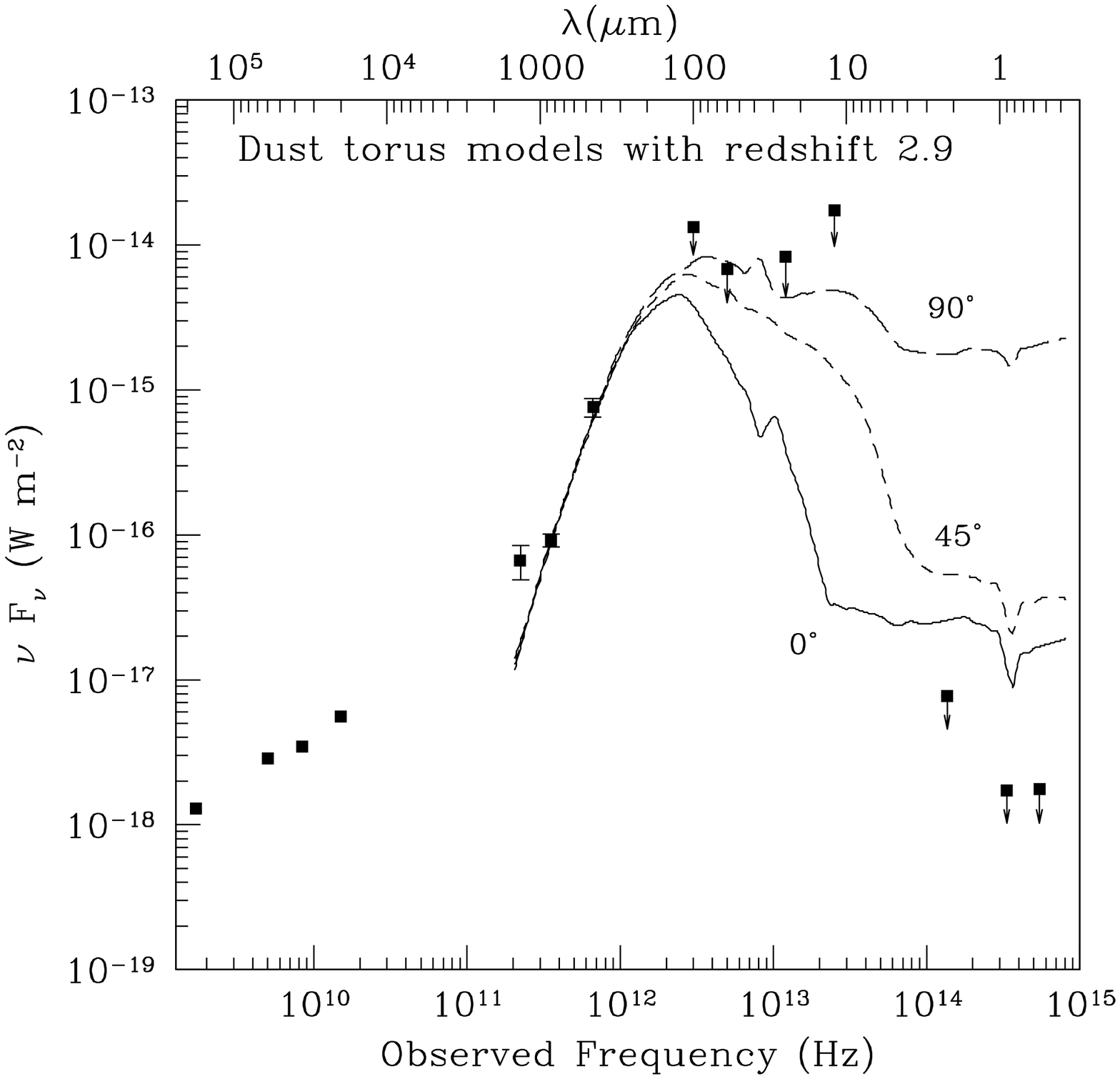}\\
\leavevmode \epsfysize=8cm \epsfbox{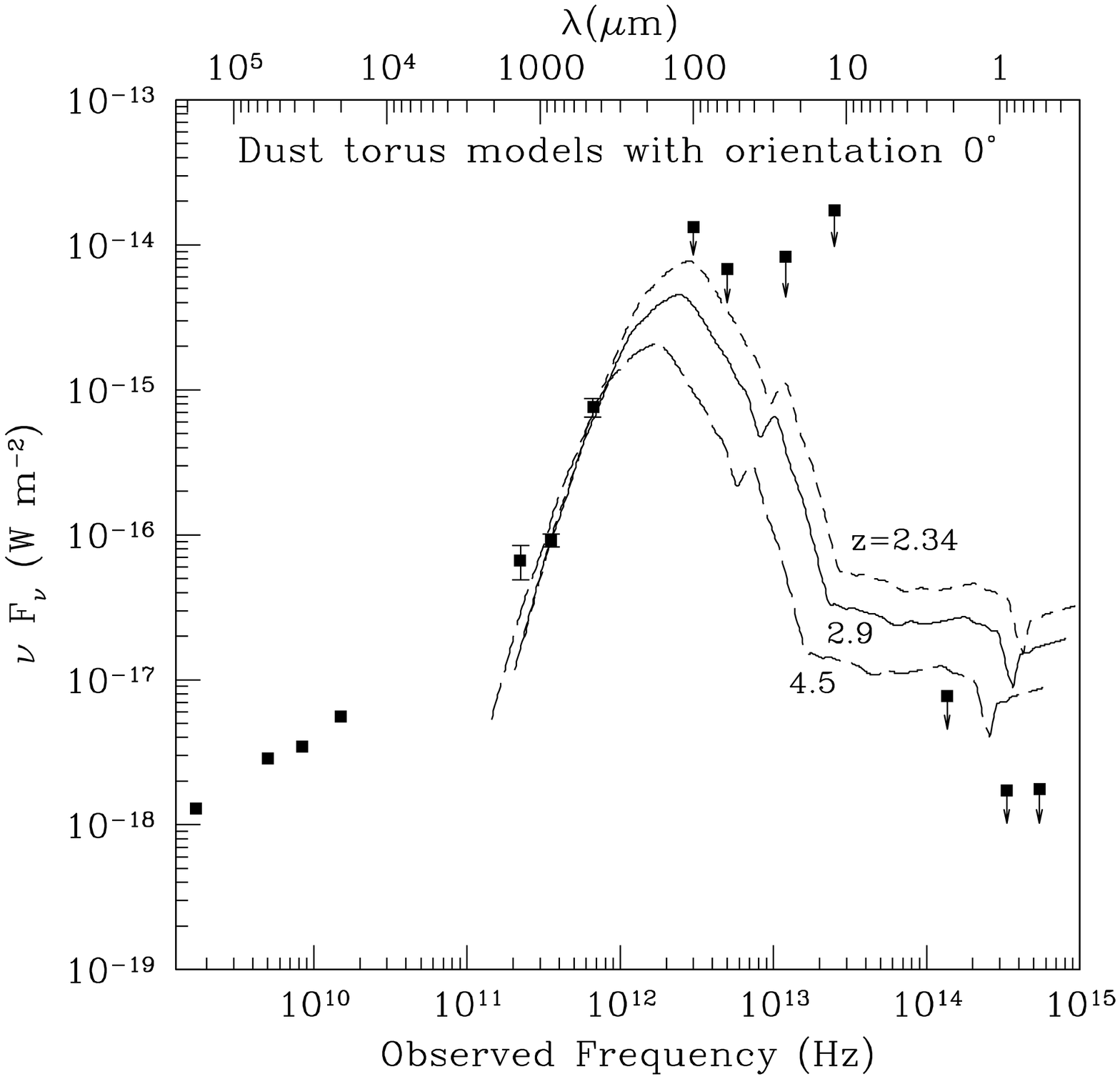}
\end{center}
\caption{The spectrum of B\,1933 from radio to optical, with an
extrapolated  radio power-law component, and a dusty AGN spectrum
(see Granato et al.~1996) fit to the sub-mm points:
(a) 3 orientations of a dusty torus surrounding the central AGN assumed
to be at $z=2.34$ here --
$0^\circ$, $45^\circ$ and $90^\circ$;
(b) the SED redshifted to $z\,{=}\,2.34$ (limit from lens modelling),
$z\,{=}\,2.9$ (best guess redshift),
and $z\,{=}\,4.5$ (highest redshift consistent with this model SED).}
\label{fig:sed}
\end{figure}

All the
available data for B\,1933 from radio through  optical are plotted as
a spectral energy distribution in Fig.~\ref{fig:sed}.
The point source limiting
magnitudes from   the  background noise in  the  $K$,  $I$ and $V$ images are
converted to flux  units using the zero  points described in Fukugita
et al.~(1995).  Stickel  et al.~(1996)  show that  in many such  galaxies, the
near-IR can  be substantially brighter   than the optical. Our  $K$-band
image with a resolution of  0.31 arcsec FWHM 
allows us to  put a $5\sigma$ limit
of  $K\,{=}\,21.4$ on  point source   detections within  the field.  This is
comparable to the limit of the HST  image of $I\,{=}\,24.2$
\cite{Syketal} given
a steep spectrum like the Stickel et al.~(1996) quasar model.
The 5$\sigma$ point source sensitivity in the $V$-band image is $V\,{=}\,25.1$

%
%

The sub-mm data points are used to constrain models consisting of
a thick dusty torus, possibly associated with an ongoing burst of
star formation, surrounding the central AGN \cite{Graetal}.
These models were shown to provide good fits to several high redshift
ultraluminous systems, thereby throwing into contention the assumption that
star formation is the dominant engine in objects such as \IRAS\,F10214+4724
(see Granato et al. 1996, 1997).
The standard  grey-body  emission from  a  dusty cloud  is
expected to dominate the sub-mm emission:
$F_\nu \propto\nu^\beta B(\nu)$.
We fix the dust  emissivity  and  temperature at
typical values for this type of object, $\beta  = 1.5$, $T = 80\,$K.
A single parameter then describes the orientation of the
dust  torus with respect  to the line  of  sight, and a nearly edge-on
view (corresponding to $0^\circ$) is required to obscure the quasar
continuum source  enough to approach the upper limits of the  $K$ and $I$
bands (see Fig.~\ref{fig:sed}a, which is for the best estimate of
the redshift, $z=2.9$).
None of the plotted curves actually falls under the near-IR limits.  However,
there is enough extra freedom in modelling the dust torus that one cannot
conclude from this that extra obscuration is required.  On the other hand
small amounts of dust in the foreground lens can certainly be invoked
to explain the lack of optical/near-IR counterparts 
\cite{Laretal,Lawetal,MalRT}. 
It is generally easier to fit the near-IR limits if the source is at
higher redshift, as shown in Fig.~\ref{fig:sed}b.
However it must be stressed that HST near-IR observations are showing this
hypothesis to be false in some objects 
\cite{Kocetal}.

\begin{figure}
\begin{center}
\leavevmode \epsfysize=8cm \epsfbox{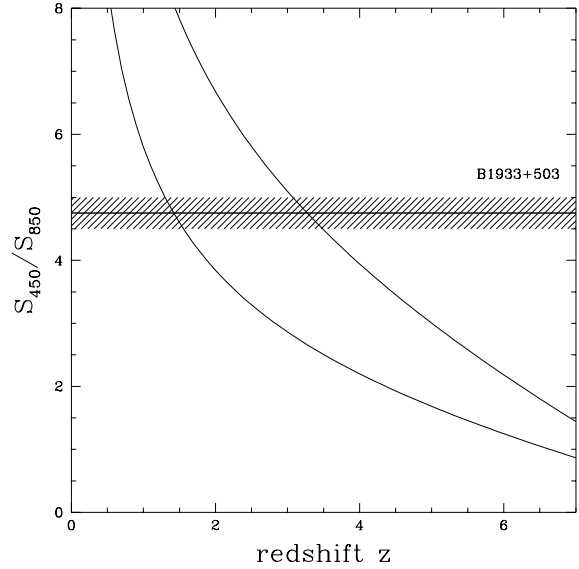}\\
\end{center}
\caption{450\mum\ to 850\mum flux ratio for different models.  The two
solid lines span a range of reasonable emissivities and dust temperatures.}
\label{fig:ratio}
\end{figure}

These model SEDs are difficult to fit to the sub-mm/FIR data 
if the redshift lies much below our estimated redshift limit (for
$\Omega_0\,{=}\,1$) of $z\,{=}\,2.34$. In Fig.~\ref{fig:ratio}
we plot the sub-mm flux ratio $S_{450}/S_{850}$ as a function of 
redshift for curves spanning 
a range of models of dusty star-forming/AGN galaxies (similar
to fig.~3 of Hughes et al.~1998), 
incorporating extreme parameters for optically thin cool dust emission 
($30\,{\rm K} < T_{\rm d} < 90\,{\rm K}$; $1 < \beta < 2$). 
We thereby place fairly robust upper and lower limits on the permissible
redshift of $1.3 < z< 3.5$. However the actual geometry and emission
characteristics of the source are unknown, and there are conceivable
scenarios where these limits might not hold (e.g.~if there were a much
larger non-thermal contribution at 850\mum\ than assumed here).
So again this argument is suggestive only, although it is now the third
such argument pointing to $z>2$.

The most statistically likely redshift is $z\,{=}\,2.9$.
However, our lens model, with reasonable parameters for the
foreground lens, is best fit with even higher values, and
with a lower limit to the
redshift of $z\,{=}\,2.34$. It is also clear from this dust torus model that 
redshifts in excess of $z\,{=}\,4.5$ can remain consistent  with the  sub-mm
and far-IR  data points, especially if the dust temperature is increased
(see Fig.~\ref{fig:sed}b). 
There is a reasonable lower limit on the redshift from our
sub-mm data, and certainly by $z\,{=}\,0.755$,
the redshift of the foreground lens, they are not even consistent with the
\IRAS\ upper limits.  From the weight of these individual redshift
arguments it seems fairly robust to assume $z>2$ for the source.

Although the 1350\mum\ point seems to imply  that some radio power law
component  is still contributing to  the emission, the results for our
dust   emission  model fit   are  not  substantially affected.  If the
non-thermal  source contributes at  850\mum\ at all, a direct extrapolation
from the radio shows it is  minimal,  
and at most would   affect the dust  emissivity
model by a factor of less than 0.2.  

We can estimate the dust mass  using the optically thin approximation
as   $1.0\times 10^8\,A_{\rm IR}^{-1}h^{-2}{\rm M_\odot}$, with
$\beta\,{=}\,1.5, T\,{=}\,80\,$K, and $A_{\rm IR}$ the 
sub-mm lens amplification (and here using $z_{\rm s}=2.9$ in an $\Omega_0=1$
universe).  If we assume that some fraction $f_{\rm SF}$ of
the sub-mm  emission is due to dust  heated by young stars,  rather than
the   central AGN  continuum source, we can estimate a 
star formation rate of
$400\times f_{\rm SF}\, A_{\rm IR}^{-1}\,h^{-2}$.  
Calzetti, Kinney \& Storchi-Bergmann~(1994) have shown that the amount of
obscuration is highly dependent on the distribution of  the dust.  The
fact that the optical/near-IR counterparts to the bright radio nucleus
are not observed can easily  be explained by a combination of
dust obscuration in
B\,1933 itself, and dust in the foreground lens.

%
%

\section{Discussion}

Given that the radio properties of B\,1933 suggest that it is a
radio-loud elliptical, it is reasonable to assume that
the  sub-mm emission  results from a roughly
edge-on  dusty disc surrounding the  core of the  galaxy and heated by
the AGN.  However, if the AGN phase 
of massive galaxies has any connection with enhanced star formation rate,
then it is
certainly possible that  some fraction of  the sub-mm  emission results
from dust heated  by young   massive stars.
The radio luminosity  is, however, not a good  measure of the AGN
sub-mm flux, and so we cannot ascertain what fraction of the dust  is
heated by star formation rather than AGN activity without a spectrum of the
source.

An HST-NICMOS  program (H.-W. Rix, private communication) will image  
B\,1933 at $H$-band  in the near future,
and may reveal the source, including some morphological information.
This should
add a strong constraint to  the lensing  predictions. Our 
work predicts that the  $H$-band  magnitude must be  greater
than 23, so that it may be within the  $5\sigma$ detection threshold of the HST
observation for a reasonable exposure time.

Although the beamsizes of the sub-mm measurements are fairly large,
substantially larger than
all  the  lensed components, the   average  SCUBA source
density has been shown to be $\sim 2000$ degree$^{-2}$ (e.g.~Smail et al.~1997,
Hughes et al.~1998, Eales et al.~1998)
at the  depth  of $\sim2.5\,$mJy RMS that  we reach  at 850\mum\ (the most
sensitive of our observational windows), and it is unlikely that another
dusty source contributes to the sub-mm flux.  We expect only one
24\,mJy source per  0.1  degree$^2$  and therefore the chance for a spurious
detection of an unrelated source is negligible.

Blain has (1996) modelled the effects  of  lensing on the distribution  of
sub-mm sources. A large  fraction of the galaxies detected at
850\mum\ flux   densities  in the  range  0.1--10\,mJy  could   be
gravitationally  lensed,
implying that lensing in the sub-mm may be a very important consideration for
high redshift source counts. Systems such  as
B\,1933 at $\sim$2\,mJy intrinsic flux density, as well as the Cloverleaf
quasar and \IRAS\,F10214+4724 \cite{Baretal}
are known lensed sub-mm sources, which
are relatively extreme objects in the amount that they are amplified.
We have  then caught a rather  rare  view of  a somewhat average dusty
radio galaxy at an early epoch, thanks to an excellent chance alignment
with  a  foreground galaxy.  This  high redshift  object appears to be
very much a scaled  down version of its more extreme cousins (such as
SMM\,02399-0136 \cite{Ivietal}, 8C1435+635 \cite{Ivietalb}, or
APM\,08279+5255 \cite{Lewetal}, which are  
otherwise the only objects bright enough to study in detail.

It is interesting to  note that if  this object were unlensed, or  only
lensed  by the  potential  of  a rich   cluster  such  as  the sources
discovered in cluster surveys (Smail et al.~1997, 1998; Chapman et al.~in
preparation), then the sub-mm emission at 850 and 450\mum\ would still be
detectable by SCUBA in a deep blank field integration,  
but the 1350\mum\ emission would be undetectable with SCUBA, and
the radio components 
would be hard to detect in a VLA integration, such as that used
for the discovery of B1933 with a 1$\sigma$ RMS of $\sim$0.4\,mJy
\cite{Syketal}. The optical identification
would then be  very difficult, given that  the true source might have
$I\,{\ge}\,27$.
There  would be no way to  assess the AGN contribution  to the
source, and the  object would likely be identified  as a star  forming
galaxy with a fairly large SFR of  $\ge 100\,{\rm M}_\odot\,{\rm yr}^{-1}$.
We hypothesize that a substantial fraction of the faint dusty  galaxies
discovered in  sub-mm surveys  could harbour  AGN as  their  main
power source, affecting the inferred comoving SFR at high $z$.
Although optical surveys have shown that high redshift
quasar systems are relatively rare, this could of course be
affected by dust obscuration, as in the case of B\,1933.
Indeed, serendipitous discovery of heavily dust-enshrouded young QSOs 
at $z\sim2$ (e.g.~H\,167, Q\,0059-2735, Egami et al.~1996)
implies a possibly sizable population of optically obscured AGN at high-$z$.

Another class of dusty objects, the EROs - extremely red objects 
with $[R-K] \ge 6$ (e.g.~Hu \& Ridgway~1994),
have also recently been detected in the sub-mm \cite{Cimetal}.
It is still unclear whether there is any AGN contribution in these objects;
B\,1933 might be another example of this class of object.

The contribution of AGN to counts of SCUBA-bright galaxies is difficult
to estimate.  This is particularly true since the volume surveyed is
essentially unknown (because the redshift range is currently unknown).
In addition there is the possible extra effect of lensing amplification.
However, the case of B\,1933 certainly leads us to be cautious in
interpretting all SCUBA detections as highly star-forming galaxies.

\section{Conclusions}

We  have  presented new  sub-mm  and
near-IR  observations    of  the complex  gravitational   lens system,
B\,1933+503.  Our estimate of the  total dust mass  is likely to be
correct to
within  the errors in our lensing  magnification estimate, since the slope
of  the  SED  shows that  optically   thin  dust  is   the most viable
explanation for the emission in this wavelength region.  The fact that
the  optical/near-IR counterparts to the bright  radio nucleus are not
observed   can easily be explained   via dust obscuration in B\,1933
itself, or by invoking a `dusty lens' (as in Lawrence et al.~1995). The large 
sub-mm flux shows that this object
is certainly not underluminous, as originally proposed.
 
The  results show  the  power of sub-millimeter observations  for
detection  and  characterization of obscured  galaxies  in the distant
universe.  Investigation of this highly amplified AGN leads us
to wonder what fraction of sources detected in `blank' SCUBA fields
could be AGN, or lensed, or even both.

\section*{ACKNOWLEDGMENTS}

This work was  supported   by the Natural Sciences   and  Engineering
Research  Council  of Canada.   The  James Clerk Maxwell  Telescope is
operated  by  The Joint  Astronomy Centre  on  behalf of  the Particle
Physics  and Astronomy Research  Council of  the  United Kingdom,  the
Netherlands Organisation for   Scientific Research,  and the  National
Research      Council      of Canada.  The Canada-France-Hawaii 
Telescope is operated by the National Research Council of Canada, 
The Centre National de la Recherche Scientifique de France and the 
University of Hawaii.
GFL is a Fellow of the Pacific Institute of Mathematical Science.
We would like to thank Chris Fassnacht for a discussion of the redshift
uncertainty from the foreground galaxy spectrum, and Charles Lawrence
for discussion of dusty lenses.

\bsp


\begin{thebibliography}{99}

\bibitem[\protect\citename{Barger et al. } 1998]{Barger} Barger A. J.,
        Cowie, L.L., Sanders D. B., Taniguchi Y., 1998, Nature, 394, 248
\bibitem[\protect\citename{Barvainis et al. }1995]{Baretal} Barvainis R.,
        Antonucci R., Hurt T., Coleman P., Reuter H.-P., 1995, ApJ, 451, L9
\bibitem[\protect\citename{Barvainis et al. }1995]{BarMAA} Barvainis R.,
        Maloney P., Antonucci R., Alloin D., 1997, ApJ, 484, 695
\bibitem[\protect\citename{Blain }1996]{Blain} Blain A.W., 1996, MNRAS, 283,
	1340
\bibitem[\protect\citename{Borys et al. }1998]{BorCS} Borys C.,
	Chapman S.C., Scott D., 1998, submitted to MNRAS, astro-ph/9808031
\bibitem[\protect\citename{Braine \& Dupraz }1994]{BraDup} Braine J.,
	Dupraz C., 1994 A\&A, 283, 407
\bibitem[\protect\citename{Calzetti et al. }1994]{Calzetti} Calzetti D.,
	Kinney, A., Storchi-Bergmann, T., 1994, ApJ, 429, 582
\bibitem[\protect\citename{Casali \& Hawarden }1992]{CasHaw} 
	Casali M., Hawarden T., 1992 JCMT-UKIRT Newsletter, 3, 33
\bibitem[\protect\citename{Cimatti et al. }1998]{Cimetal} Cimatti A.,
	Andreani P., R{\" o}ttgering H.J.A., Tilanus R., 1998, Nat, 392, 895 
\bibitem[\protect\citename{Cowie et al. }1994]{Cowetal} Cowie L.L.,
	Gardner J.P., Hu E.M., Songaila A., Hodapp K.-W., Wainscoat R.J.,
	1994, ApJ, 434, 114
\bibitem[\protect\citename{Dunlop et al. }1994]{Dunetal} Dunlop J.S.,
        Hughes D.H., Rawlings S., Eales S.A., Ward M.J., 1994, Nat, 370, 347
\bibitem[\protect\citename{Eales \& Edmunds }1996]{EalEdm} Eales S.,
	Edmunds, 1996, MNRAS, 280, 1167
\bibitem[\protect\citename{Eales et al. }1998]{Ealetal} Eales S., Lilly S.,
	Gear W., Dunne L., Bond J.R., Hammer, F., Le Fevre O., Crampton D.,
	MNRAS, submitted
\bibitem[\protect\citename{Egami et al. }1997]{Egami} Egami E., Iwamuro F.,
	Maihara T., Oya S., Cowie L.L., 1996, AJ, 112, 73
\bibitem[\protect\citename{Fukugita et al. }1995]{Fuketal} Fugugita M. et al.,
	Shimasaku K., Ichikawa T., 1995, PASP, 107, 945
\bibitem[\protect\citename{Granato et al. }1996]{Graetal} Granato G.L.,
	Danese, L., Franceschini, A., 1996, ApJL, 460, 11	
\bibitem[\protect\citename{Granato et al. }1997]{Graetalb} Granato G.L.,
	Danese, L., Franceschini, A., 1997, ApJ, 486, 147
\bibitem[\protect\citename{Holland et al. }1998]{Holetal} Holland W.S.,
	et al., 1998, MNRAS, in press
\bibitem[\protect\citename{Hu \& Ridgway }1994]{HuR} Hu E.M., Ridgway S.E.,
	1994, AJ, 107, 1303
\bibitem[\protect\citename{Hughes et al. }1997]{HugDR} Hughes D.H., 
	Dunlop J.S., Rawlings S., 1997, MNRAS, 289, 766
\bibitem[\protect\citename{Hughes et al. }1998]{Hugetal} Hughes D.H., et al.,
        1998, Nature, 394, 241
\bibitem[\protect\citename{Hutchings et al. }1998]{Hutchings} 
	Hutchings J., Crampton D., Morris S., Steinbring E. 1998, PASP, 110, 374
\bibitem[\protect\citename{Jaffe et al. }1998]{Jaffe} Jaffe W., et al.,
	1996, ApJ, 460, 214
\bibitem[\protect\citename{Ivison et al. }1998]{Ivietalb} Ivison et al., 
	1998, ApJ, 494, 211
\bibitem[\protect\citename{Ivison et al. }1998]{Ivietal} Ivison et al., 
	1998, MNRAS, 298, 583
\bibitem[\protect\citename{Jackson et al. }1995]{Jacetal} Jackson N., 1995,
	MNRAS, 274, L25
\bibitem[\protect\citename{Jenness \& Lightfoot }1998]{SURF} Jenness T.
        \& Lightfoot, J. F., 1998, in
        Albrecht, R., Hook, R.N., Bushouse, H.A., eds, ASP Conf. Ser.
        Vol.~145, Astronomical Data Analysis Systems and Software.
        Astron. Soc. Pac., San Francisco, p.$\,216$
\bibitem[\protect\citename{Kochanek }1993]{Kochanek} 
	Kochanek C.S., 1993, MNRAS, 261, 453
\bibitem[\protect\citename{Kochanek et al. }1989]{KocBLN} Kochanek C.S.,
	Blandford R.D., Lawrence C.R., Narayan R., 1989, MNRAS, 238, 43
\bibitem[\protect\citename{Kochanek }1998]{Kocetal} Kochanek C.S., Falco E.E.,
	Impey C.D., Lehar J., McLeod B.A., Rix H.-W., Keeton C.R., Peng C.Y.,
	et al., 1998, submitted to ApJ, astro-ph/9809371
\bibitem[\protect\citename{Larkin et al. }1994]{Laretal} Larkin J.E. et al.,
	1994, ApJL, 420, 9
\bibitem[\protect\citename{Lawrence et al. }1995]{Lawetal} Lawrence C. R., 
	Elston R., Januzzi B.T., Turner E.L., 1995, AJ, 110, 2570
\bibitem[\protect\citename{Lewis et al. }1998]{Lewetal} Lewis G.F.,
	Chapman S.C., Ibata R.A., Irwin M.J., Trotten E.J., 1998,
	ApJL, in press, astro-ph/9807293
\bibitem[\protect\citename{Lilly et al. }1996]{Lilly} 
	Lilly S., Le Fevre O., Hammer F., Crampton D. 1996 ApJL, 460, 1
\bibitem[\protect\citename{Malhotra et al. }1996]{MalRT} Malhotra S.,
	Rhoads J.E., Turner E.L., 1996, MNRAS, 288, 138
\bibitem[\protect\citename{Nair }1998]{Nair} Nair S., 1998, MNRAS, in press
\bibitem[\protect\citename{Pozzetti et al. }1996]{Pozzetti} Pozzetti L.,
	Bruzual A.G., Zamorani, G., 1996, MNRAS, 281, 953
\bibitem[\protect\citename{Rigaut et al. }1998]{AOB} Rigaut F. et al., 1998,
	PASP, 110, 152
\bibitem[\protect\citename{Sanders et al. }1989]{Sanetal} Sanders D., Phinney E.
	et al., 1989, ApJ, 347, 29
\bibitem[\protect\citename{Schneider }1992]{Schneider} Schneider P., Weiss A., 
	1992, A\&A, 260, 1
\bibitem[\protect\citename{Smail et al. }1997]{SmaIB} Smail I., Ivison R.J.,
        Blain A.\,W., 1997, ApJ, 490, L5 (SIB){}{}
\bibitem[\protect\citename{Smail et al. }1998]{Smaetal} Smail I., Ivison R.J.,
	Blain A.\,W., 1998, ApJL, in press 
\bibitem[\protect\citename{Stickel et al. }1996]{Stietal} Stickel M., Rieke G.,
	Kuhr H., Rieke M., 1996, ApJ, 468, 556
\bibitem[\protect\citename{Sykes et al. }1998]{Syketal} Sykes C.M. et al.,
        1998, MNRAS, in press, astro-ph/9710358
\bibitem[\protect\citename{van der Marel et al. }1996]{vandermarel} 
	van der Marel, R.P. 1991, MNRAS, 253, 710
\end{thebibliography}
\end{document}